\documentclass[aps,amsmath,amssymb,nofootinbib,superscriptaddress,showpacs,floatfix,prl,twocolumn]{revtex4-1}

\usepackage{latexsym}
\usepackage{graphicx}
\usepackage{times,psfrag,subfigure}
\usepackage{amsmath}
\usepackage{dsfont}
\usepackage{dcolumn}
\usepackage{bm,bbm}       
\usepackage{color}
\usepackage{latexsym,amsmath,amssymb,bm,euscript}
\usepackage{dsfont}
\usepackage{textcomp}

\newcommand{\beq}{\begin{equation}}
\newcommand{\eeq}{\end{equation}}
\newcommand{\beqarray}{\begin{eqnarray}}
\newcommand{\eeqarray}{\end{eqnarray}}

\begin{document}

\allowdisplaybreaks

\title{Sputtering induced re-emergence of the topological surface state in Bi$_2$Se$_3$}

\date{\today}

\author{Raquel Queiroz}
\email{r.queiroz@fkf.mpg.de}
\affiliation{Max-Planck-Institut f\"ur Festk\"orperforschung, Heisenbergstrasse 1, D-70569 Stuttgart, Germany} 
\author{Gabriel Landolt}
\affiliation{Physik-Institut, Universit\"at Z\"urich, Winterthurerstrasse 190, CH-8057 Z\"urich, Switzerland} 
\affiliation{Swiss Light Source, Paul Scherrer Institut, CH-5232 Villigen, Switzerland}
\author{Stefan Muff}
\affiliation{Institute of Condensed Matter Physics, Ecole Polytechnique F{\'e}d{\'e}rale de Lausanne, CH-1015 Lausanne, Switzerland} 
\affiliation{Swiss Light Source, Paul Scherrer Institut, CH-5232 Villigen, Switzerland}
\author{Bartosz Slomski}
\affiliation{Physik-Institut, Universit\"at Z\"urich, Winterthurerstrasse 190, CH-8057 Z\"urich, Switzerland} 
\affiliation{Swiss Light Source, Paul Scherrer Institut, CH-5232 Villigen, Switzerland}
\author{Thorsten Schmitt}
\affiliation{Swiss Light Source, Paul Scherrer Institut, CH-5232 Villigen, Switzerland}
\author{Vladimir N. Strocov}
\affiliation{Swiss Light Source, Paul Scherrer Institut, CH-5232 Villigen, Switzerland}
\author{Jianli Mi}
\affiliation{Department of Physics and Astronomy, Interdisciplinary Nanoscience Center, Aarhus University, 8000 Aarhus C,      Denmark}
\author{Bo Brummerstedt Iversen}
\affiliation{Department of Physics and Astronomy, Interdisciplinary Nanoscience Center, Aarhus University, 8000 Aarhus C,      Denmark}
\author{Philip Hofmann}
\affiliation{Department of Physics and Astronomy, Interdisciplinary Nanoscience Center, Aarhus University, 8000 Aarhus C,      Denmark}
\author{ J\"urg Osterwalder}
\affiliation{Physik-Institut, Universit\"at Z\"urich, Winterthurerstrasse 190, CH-8057 Z\"urich, Switzerland} 
\author{Andreas P. Schnyder}
\email{a.schnyder@fkf.mpg.de}
\affiliation{Max-Planck-Institut f\"ur Festk\"orperforschung, Heisenbergstrasse 1, D-70569 Stuttgart, Germany} 
\author{J. Hugo Dil}
\email{hugo.dil@epfl.ch}
\affiliation{Institute of Condensed Matter Physics, Ecole Polytechnique F{\'e}d{\'e}rale de Lausanne, CH-1015 Lausanne, Switzerland} 
\affiliation{Swiss Light Source, Paul Scherrer Institut, CH-5232 Villigen, Switzerland}
  
\begin{abstract}
We study the fate of the surface states of Bi$_2$Se$_3$ under disorder with strength larger than the bulk gap, 
caused by neon
sputtering and nonmagnetic adsorbates. We find that neon sputtering introduces strong but dilute defects, which can be modeled by a unitary impurity distribution,
whereas adsorbates, such as water vapor or carbon monoxide, are best described by   Gaussian disorder. 
Remarkably, these two disorder types have a dramatically different effect on the surface states.
Our soft x-ray ARPES measurements combined with numerical simulations show that
unitary surface disorder pushes the Dirac state to inward  quintuplet layers, burying it below
an insulating surface layer. As a consequence, the surface spectral function becomes weaker, but retains its 
quasiparticle peak. This is in contrast to Gaussian disorder, which
 smears out the quasiparticle peak completely.
 At the surface of Bi$_2$Se$_3$, the effects of Gaussian
disorder can be reduced by removing surface adsorbates using neon sputtering,
which, however, introduces unitary scatterers.
Since unitary disorder has a weaker effect than Gaussian disorder, the ARPES signal of the Dirac surface state becomes sharper upon sputtering. 
\end{abstract}

\date{\today}



\maketitle

\emph{Introduction.--}\label{Sec:Theo}
An important hallmark of three-dimensional topological insulators are their protected Dirac-cone surface states,
which connect bulk valence and conduction bands~\cite{qi:rmp,hasan:rmp,chiu_review15}. Since these
surface states arise due to a nontrivial wave funtion topology in the bulk~\cite{Kane2005},
they are robust to nonmagnetic disorder
with strength $\gamma$ smaller than the bulk gap $\Delta$~\cite{nomuraPRL07,ryuPRL07,ostrovskyPRL07,ryuNJP10,Schnyder2008,kim_fran_refael_PRB_15,essin_gurarie_15}. 
Moreover, their existence is independent of the surface orientation~\cite{Moon2011,Teo2008} and the local
surface chemistry.  Protected topological surface states 
have been experimentally
observed in numerous topological insulators, such as Bi$_{1-x}$Sb$_x$~\cite{Hsieh2008}, Bi$_2$Se$_3$~\cite{Hsieh2009}, Bi$_2$Te$_3$~\cite{hsiehPRL09}, by both
angle-resolved photoemission spectroscopy (ARPES)~\cite{Chen2009,Xia2009,Souma2011,Sato2011,beniaPRB13,bahramyNatCommun12}
and scanning tunneling experiments~\cite{roushanNature09,SeoNature10}.
State of the art spin-resolved ARPES has allowed to map out the predicted
helical spin texture of the surface states~\cite{Hsieh2009-2,Xu2011,Xu2012,Eremeev2012,Fu2009}. However, in recent experiments only little attention has been given to the fundamental property that gave these materials their name, namely the topological protection of the surface states
against disorder~\cite{alpichshevPRL12,pielmeier2015}.

The surface of Bi$_2$Se$_3$ and other Bi-based topological insulators, adsorbs H$_2$O, H$_2$, or CO molecules upon exposure to air
or vacuum rest gas~\cite{Bianchi2010,Bianchi2011,beniaPRL11}.
This introduces a large number of 
 impurity scatterers.  Furthermore, as a result of chemical bonding between the adsorbates and the surface,
the number of surface carriers is increased, which leads  to 
band bending and the
development of two-dimensional surface quantum well states~\cite{Bianchi2010,Bianchi2011}. 
Additional scattering centers on 
 Bi$_2$Se$_3$ surfaces are caused by
step edges and Se vacancies, which host 
impurity bound states with energies of the order or
larger than the band gap~\cite{Hatch2011,alpichshevPRL12}.
While the adsorbates lead to a dense distribution of relatively weak impurities, 
the step edges and Se vacancies introduce a dilute distribution 
of very strong scatterers. 
The former is commonly called Gaussian disorder~\cite{atkinsonPRL00,chamonPRB01,QueirozPRB14,Queiroz2015} and the latter is known as unitary disorder. 
Unitary scatterers can also be  artificially created on the surface by sputtering it 
with neon or argon ions~\cite{Hatch2011}.
Due to their topological protection, the surface states are robust against 
both types of disorder as long as the disorder strength $\gamma$ is smaller than the bulk gap $\Delta$~\cite{nomuraPRL07,ryuPRL07,ostrovskyPRL07,ryuNJP10,Schubert2012,QueirozPRB14,Queiroz2015}, but not necessarily otherwise.

\begin{figure}[t!]
\includegraphics[width=0.8\columnwidth]{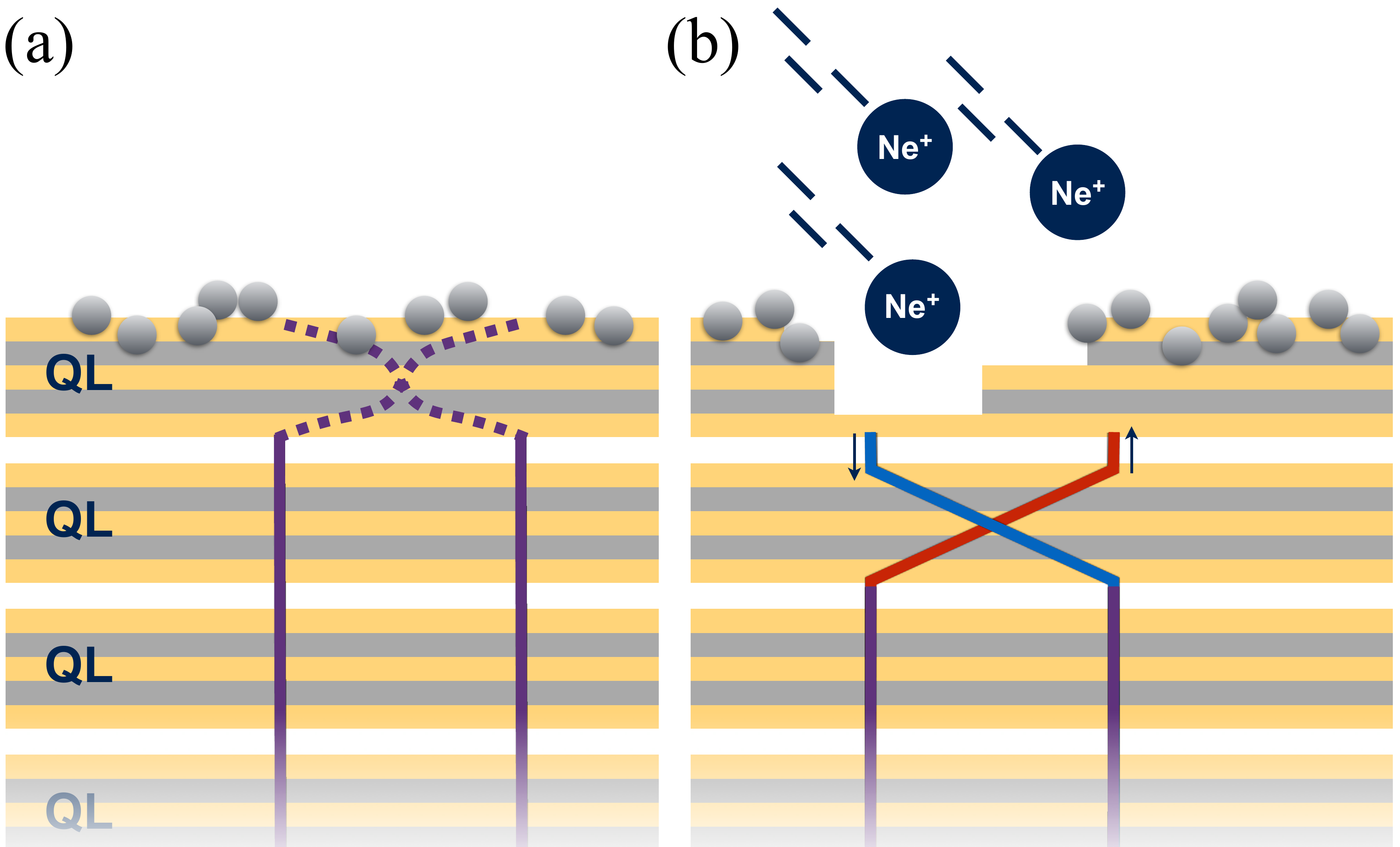}
\caption{ \label{Fig1}
{(color online). Illustration of the re-emergence of the topological surface state in Bi$_2$Se$_3$ upon sputtering. 
(a) Vacuum rest gas exposure leads to the absorption of water vapor and other molecules, giving rise
to Gaussian surface disorder (grey circles). This results in a broadening and weakening of the ARPES spectral function.
(b) Neon sputtering introduces vacancies and defects at the surface, thereby pushing the surface states
to inward { quintuplet layers (QL).} }
}
\end{figure}

\begin{figure*}[t!]
\centering
\includegraphics[clip,angle=0,width=0.9\textwidth]{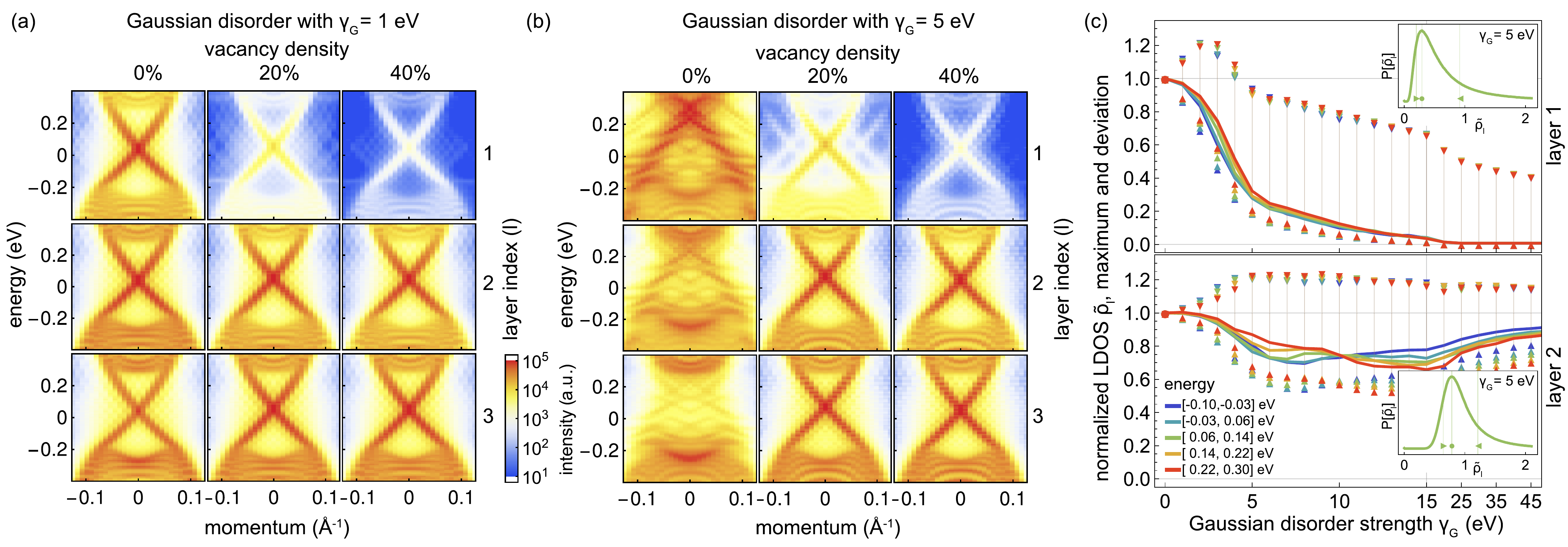}
\caption{(color online). \label{mFig2}
(a), (b): Layer resolved spectral function $ A_l (\omega,{\bf k}_{\parallel} )$, Eq.~\eqref{Eq:intensity}, as a function of surface momentum ${\bf k}_{\parallel}$ along  the $\overline{\textrm M}$--$\overline{\Gamma}$--$\overline{\textrm M}$ direction
of the (001) surface Brillouin zone, obtained by diagonalizing tight-binding model~\eqref{Eq:Hamiltonian}.
The magnitude of $ A_l (\omega,{\bf k}_{\parallel} )$ is indicated by the { logarithmic} color scale, with blue and red representing low and 
high intensity, respectively. { The same color scale is used for all subpanels.}
The effects of surface adsorbates is simulated by Gaussian disorder with (a) $\gamma_{\textrm{G}}= 1$~eV
and (b)  $\gamma_{\textrm{G}}= 5$~eV. 
To mimic the sputtering process, the density of surface defects is increased from $\gamma_{\textrm{U}}$ = 0\% 
in the left columns to $\gamma_{\textrm{U}}$ = 40\% in the right columns.
(c) 
{ Energy-resolved distribution of the local density of
states $P [ \tilde{\rho}_l (\omega) ]$ for the surface and the first inward { quintuplet} layer as a function of Gaussian disorder strength.
The solid line represents the maximum of the distribution, whereas the triangles and inverted triangles indicate
the left and right standard deviation, respectively~\cite{supplement}.  
The insets show the probability distributions for the disorder strength $\gamma_{\textrm G}=5$~eV.
}}
\label{fig:Fig2}
\end{figure*}

\begin{figure*}[t]
\begin{center}
\includegraphics[width=0.76\textwidth]{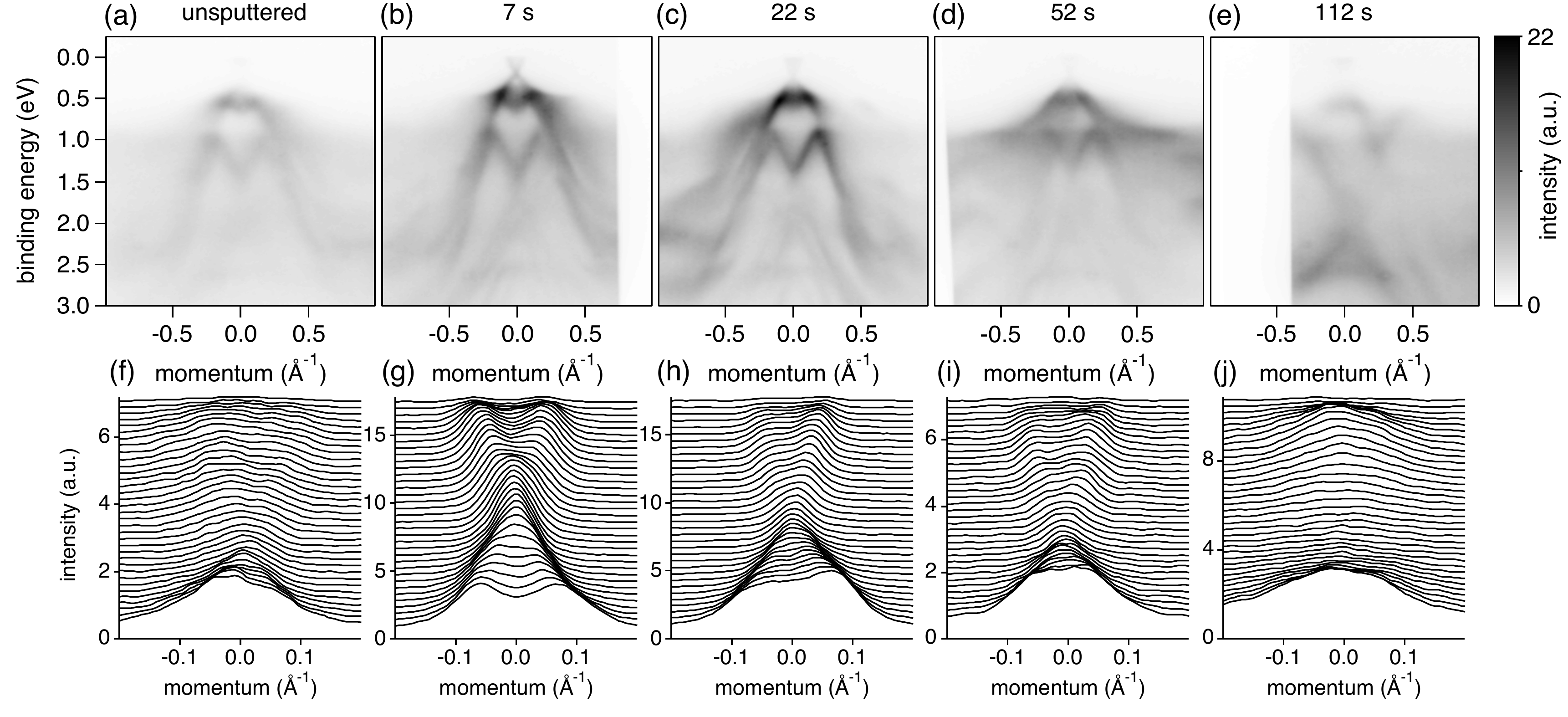}
\caption{
(color online). Effects of  neon sputtering on the (001) surface states of Bi$_2$Se$_3$.
The upper row [panels (a)-(e)] shows SX-ARPES band maps with a photon energy of $h \nu =$ 380 eV
along the $\overline{\textrm M}$--$\overline{\Gamma}$--$\overline{\textrm M}$ crystallographic direction
and as a function of neon sputtering.
In all panels the same grey scale is used to indicated the ARPES intensity, with 
white and black corresponding to low and high signal, respectively.
 The lower row [panels (f)-(j)] displays the momentum distribution curves (MDCs) that correspond 
 to the band maps of the upper row.
Each curve is shifted vertically for clarity. 
The MDCs from the top to the bottom have binding energies from 0~meV up to 442~meV
 with a spacing of 13~meV between traces.
 The SX-ARPES band maps and momentum distribution curves 
 with a photon energy of $h \nu =725$~eV 
 are provided in Fig.~\ref{SFig1} of the SM \cite{supplement}.
}
\label{fig:AllData}
\label{mFig3}
\end{center}
\end{figure*}

 In this Letter, we combine soft x-ray ARPES with numerical simulations to study how
the  surface states of Bi$_2$Se$_3$ are modified in the presence of strong disorder with $\gamma \gg \Delta$.
We use UHV rest gas exposure to create Gaussian  disorder and employ neon sputtering to introduce unitary disorder on the surface of Bi$_2$Se$_3$ { (Fig.~\ref{Fig1})}.
Our numerical simulations show that the type of disorder matters for the topological insulator surface state.
We find that Gaussian disorder with $\gamma \gg \Delta$ introduces a large number of impurity scatterers,
which leads to a strong coupling between the bulk and surface states~\cite{Schubert2012}.
As a consequence, the momentum-space structure of the Dirac surface state is completely
smeared out, leading to a surface spectral function that only exhibits broad features but no
sharp quasiparticle peak (Fig.~\ref{fig:Fig2}). 
Unitary disorder, on the other hand, creates a topologically trivial insulator at the surface,
thereby pushing the Dirac state to inward { quintuplet} layers which are less disordered.
Hence, unitary scatterers do not destroy the quasiparticle peak in the surface spectral function, but only
reduce its sharpness and intensity.
Our ARPES measurements of Bi$_2$Se$_3$ surfaces show that
the spectral function becomes sharper upon sputtering (Fig.~\ref{fig:AllData}). This seemingly paradoxical observation
is explained by taking into account the different disorder types of the surface adsorbates
and the sputtering-induced impurities.
That is, neon sputtering reduces the effects of Gaussian disorder by removing the surface adsorbates,
which however comes at the expense of introducing unitary scatterers.
Since unitary disorder has a weaker effect on the spectral function than Gaussian one,
the quasipartilce peaks in the ARPES signal 
 become more pronounced due to sputtering.

\emph{Numerical simulations.--}\label{Sec:Theo}
To simulate the effects of disorder on the  Bi$_2$Se$_3$ surface states,
we employ a low-energy tight-binding Hamiltonian that describes the Bi-$p_z$ and Se-$p_z$ orbital bands close
to the $\Gamma$ point of the Brillouin zone~\cite{Zhang2009}. 
The Hamiltonian can be conveniently expressed in terms of the spinor
$\Phi_{\bf k}=(|p_{z,{\bf k}}^1,\uparrow\rangle,|p_{z,{\bf k}}^1,\downarrow\rangle,|p_{z,{\bf k}}^2,\uparrow\rangle,|p_{z,{\bf k}}^2,\downarrow\rangle)$
as 
$\mathcal{H}=\frac{1}{2}\sum\limits_{\bf k}\Phi^\dag_{\bf k}H^{\phantom{\dag}}_{\bf k}\Phi^{\phantom{\dag}}_{\bf k}$, with 
\begin{equation} \label{Eq:Hamiltonian} 
H_{\bf k}=\epsilon_{\bf k}\sigma_0\otimes\tau_0+m_{\bf k}\sigma_0\otimes\tau_3+\sum_{i=0}^2a^i_{\bf k}\sigma_i\otimes\tau_1,
\end{equation}
where the two sets of Pauli matrices $\sigma_{\alpha}$ and $\tau_\alpha$ describe 
the spin and orbital degrees of freedom, respectively.
The tight-binding model $H_{\bf k}$ is defined on a rhombohedral lattice with lattice constants $a=4.08$ \AA \,
and $c=29.8$ \AA .
Eq.~\eqref{Eq:Hamiltonian}
contains a kinetic term $\epsilon_{\bf k}=D_1[1- \cos (k_z c)]+D_2[3-2\cos (k_x\sqrt{3} a/2 ) \cos (k_ya/2)-\cos (k_y a)]-\mu $, a mass term $m_{\bf k}=B_1[1 - \cos (k_z c)]+B_2[3-2\cos (k_x\sqrt{3} a/2 ) \cos (k_ya/2)-\cos ( k_y a) ]+M$, and an interorbital
coupling, which is parameterized by the vector ${\bf a}_{\bf k}$ with the three components
$a^0_{\bf k}=A_0\sin(k_z c)$, $a^1_{\bf k}=A_1\sqrt{3} \sin (	k_x \sqrt{3} a/2) \cos (k_y a/2)$, and $a^2_{\bf k}=A_1 [ \cos (	k_x \sqrt{3} a/2) \sin (k_y a/2) + \sin(k_y a)]$.
For the numerical calculations, the tight-binding parameters  are
determined by a fit of the energy spectrum of $H_{\bf k}$ to that
of ab inito DFT calculations~ \cite{Zhang2009,sasakiAndoPRL11,parameters}.
We observe that Hamiltonian~\eqref{Eq:Hamiltonian} satisfies time-reversal symmetry,
but importantly breaks sublattice (chiral) symmetry~\cite{essin_gurarie_15},
which is in contrast to the model considered in Ref.~\cite{Schubert2012}.

We implement surface disorder due to adsorbates and lattice defects by adding the  potential  $\delta\mathcal{H}$
to the Hamiltonian~\eqref{Eq:Hamiltonian}, with
\begin{equation} \label{Eq:DisorderPotential}
\delta\mathcal{H}=\sum_{\bf k,q_\parallel}\sum_{a=\textrm{G},\textrm{U}}\Phi^\dag_{\bf k}
\sigma_0\otimes [ V^{\textrm{m}}_a({\bf q_\parallel})\tau_3 + V^\mu_a({\bf q_\parallel})\tau_0 ] \Phi^{\phantom{\dag}}_{\bf k+q_\parallel},
\end{equation}
where  $V^{b}_{a}({\bf q_\parallel}) = (1 / \sqrt{\mathcal{N}} ) \sum_n v^{b}_{a} ( {\bf r}_n ) e^{- i {\bf q}_{\parallel} \cdot {\bf r}_n }$ represents the Fourier transform of the uncorrelated random onsite potentials $v^{b}_{a} ( {\bf r}_n )$
at the surface sites $ {\bf r}_n$.
The disorder potential $\delta H$ includes both  local variations in the mass term ($b=\textrm{m}$)
and in the chemical potential ($b=\mu$). For the surface adsorbates
we employ a Gaussian type disorder distribution ($a=\textrm{G}$), whereas for the lattice defects 
a unitary disorder distribution  ($a=\textrm{U}$) is used.
To realize the latter, we randomly choose $N_{\textrm{imp}}$ impurity sites ${\bf r}_i$ ($i = 1, 2, \ldots, N_{\textrm{imp}}$) at the surface,
which all have the same large onsite  potential $v^b_{\textrm{U}}({\bf r}_i) = 10^5$~eV.
To implement the former,  random potentials $v^{b}_{\textrm{G}} ( {\bf r}_n )$  at  each lattice site  ${\bf r}_n$
are drawn from a box distribution with width $\gamma_{\textrm{G}}$ and $p [v^b_{\textrm{G}} ({\bf r}_n )] = 1/ \gamma_{\textrm{G}}$ for $v^b_{\textrm{G}}({\bf r}_n ) \in [-\gamma_{\textrm{G}}/2, + \gamma_{\textrm{G}} /2]$~\cite{atkinsonPRL00}.
Note that the strength of the Gaussian disorder is determined by the width $\gamma_{\mathrm{G}}$ of the distribution, 
while the strength of the unitary disorder is determined by
the  impurity density $\gamma_{\textrm{U}}=N_\textrm{imp}/N_\textrm{tot}$, where
$N_\textrm{tot}$ is the total number of surface sites.

The effects of impurities on the Dirac surface state are best revealed by examining 
the momentum-resolved spectral function $A_l ( \omega, {\bf k}_\parallel )$~\cite{Schubert2012,Queiroz2015}, which is directly proportional to the ARPES intensity.
The spectral function in the $l$th { quintuplet} layer is given by
\begin{equation} \label{Eq:intensity}
 A_l (\omega,{\bf k}_{\parallel} )=
-\frac{\hbar}{4 \pi}\text{Im}\sum\limits_{m,\nu}\frac{\left| \frac{1}{\sqrt{\mathcal{N}}} \sum_{n}\Psi^m_{\nu,l}({\bf r}_n)e^{-i {\bf  k}_\parallel \cdot {\bf r}_n }\right|^2}{\omega-E_m +i \eta},
\end{equation}
where $\nu$ is the band index and ($E_m$, $\Psi^m_{\nu, l}$) represents the eigensystem of Hamiltonian~\eqref{Eq:Hamiltonian}
with surface disorder~\eqref{Eq:DisorderPotential}.
Using exact diagonalization algorithms~\cite{ARPACK} we evaluate 
expression~\eqref{Eq:intensity} for a (001) slab of dimension $100 \times 25 \times 25$
and an intrinsic broadening of $\eta=0.02$ eV.

Figure~\ref{mFig2} shows the spectral function for the first three outermost layers in the presence of
unitary scatterers
and Gaussian disorder with  strength $\gamma_{\textrm{G}}=1$~eV in panel (a) and 
$\gamma_{\textrm{G}}=5$~eV in panel (b).
We observe that Gaussian disorder with $\gamma_{\textrm{G}}=1$~eV, 
which is of the order of the bulk gap $\Delta$,  does not alter the surface states, apart
from small broadening effects.
That is, the Dirac cone surface state is clearly visible as a sharp quasiparticle peak
in $ A_l (\omega,{\bf k}_{\parallel} )$, which decays
exponentially into the bulk 
on the length scale of about three { quintuplet} layers [left column in Fig.~\ref{mFig2}(a)]. 
However, for Gaussian disorder with $\gamma_{\textrm{G}}=5$~eV,  
the momentum-space
structure of the Dirac state is completely destroyed.
This is due to a large number of impurity bound states at the surface, which 
leads to a strong coupling between bulk and surface states.
Thus,  $A_l (\omega,{\bf k}_{\parallel} )$ in the presence of very strong Gaussian 
disorder exhibits only broad humps,
but no sharp features  [left column in Fig.~\ref{mFig2}(b)].
It is important to note that even though the disorder $\delta \mathcal{H}$
completely breaks translation symmetry along the surface,  
the $\mathbb{Z}_2$ topological invariant in the bulk remains
well defined. 
As a consequence, there exists a band of delocalized sub-gap states at the boundary 
of the topological insulator $H_{\bf k}$, even for very strong surface disorder~\cite{Schnyder2008,ryuNJP10,ostrovskyPRL07,kim_fran_refael_PRB_15,essin_gurarie_15}. 
Since these delocalized states are not eigenstates of the surface momentum, 
they do not reveal themselves in 
Fig.~\ref{mFig2}(b).
{ However, the { distribution of the local density of states $P[\tilde{\rho}_l]$  }
 for $\gamma_{\textrm{G}}=5$~eV indicates that there exist delocalized states
 both in the surface layer ($l=1$) and the first inward { quintuplet} layer ($l=2$) [Fig.~\ref{fig:Fig2}(c)] { (for details see Ref.~\cite{supplement})}.
That is, $P[\tilde{\rho}_{l=2}]$ is peaked close to $\tilde{\rho}_l = 1$,
while $P[\tilde{\rho}_{l=1}]$ 
 exhibits a tail that extends beyond $\tilde{\rho}_l = 1$~\cite{supplement,schubert_vollhardt_PRB_10}.
In fact, our numerical data show that the surface state can never be completely localized.
Even for arbitrarily strong Gaussian disorder, $P[\tilde{\rho}_{l=2}]$ has its
maximum close to $\tilde{\rho}_l = 1$.
We note that $P[\tilde{\rho}_l]$ is almost energy independent, which
is in contrast to the sublattice symmetric model of Ref.~\cite{Schubert2012}, 
where $\omega = 0$ is distinct from  other energies. }

To simulate the sputtering process, we increase the density of surface defects
from $\gamma_{\textrm{U}} = 0\%$ to  $\gamma_{\textrm{U}} = 40\%$.
The right columns of Figs.~\ref{mFig2}(a) and~\ref{mFig2}(b) show that
for $\gamma_{\textrm{U}} = 40$\% the Dirac state is pushed to 
the second and third inward { quintuplet} layers, since the surface layer ($l=1$) becomes more and more 
insulating.
As a result the Dirac state reappears in the spectral function of the
second and third layers as well-defined quasiparticle peaks.

\emph{SX-ARPES measurements.--}
\label{Sec:Exp}
In order to measure the effects of sputtering on the Dirac surface state using ARPES,
it is  necessary to employ large incident photon energies, such that the probing depth is considerably larger
than one  Bi$_2$Se$_3$ unit cell. 
Therefore, instead of conventional UV-ARPES,
we used soft x-ray (SX) ARPES with a probing depth, defined as 3 mean free path lengths, of about 30~\AA~\cite{strocov2010,Powell19991}, which
corresponds to three Bi$_2$Se$_3$ quintuplet layers.
The photoemission experiments were performed on in-situ cleaved single crystals of Bi$_2$Se$_3$ at the ADRESS beamline
of the Swiss Light Source  using photon energies of $h \nu =$ 380~eV and $h \nu =$ 725~eV.
To prevent freezing of the sputtering agent, the  samples were cleaved and sputtered at room temperature. 
The SX-ARPES measurement, however, was carried out at the low temperature of 10 K, since otherwise
the photoemission signal would be too blurry, due to the loss of spectral coherence as expressed by the large Debye-Waller factor~\cite{Hofmann2002,Braun:2013}.

Our aim was to investigate 
how the surface spectral functions is changed
as the density of unitary scatterers is increased, while the 
Gaussian disorder of the surface adsorbates is kept constant. 
For that purpose, it was crucial to keep the times between 
cleaving, sputtering, cooling, and measuring the crystals fixed, such that all samples are exposed 
for the same duration to the UHV rest
gas at a base pressure better than $5\times 10^{-10}$~mbar. The time between cooling down the sample and measuring was of the order of half an hour, 
resulting in the
adsorbtion of a significant amount of H$_2$, CO, and H$_2$O molecules.  
Different densities of unitary disorder were introduced by
irradiating the Bi$_2$Se$_3$ surface for different time periods with
Ne$^+$ ions with an energy of 0.7 keV at a pressure of $3\cdot10^{-6}$ mbar. For every sputtering cycle a new sample was cleaved under identical circumstances. 
After sputtering, the sample was immediately in-situ transferred to the 
measurement chamber to start the SX-ARPES experiment. To assure the reproducibility of the results the sample surface was scanned for homogeneity and the measurements, especially for the unsputtered and lightly sputtered samples, were repeated several times. All data shown here was obtained under identical conditions with respect to beamline and analyser settings, and integration time.

In Fig.~\ref{fig:AllData} we present the SX-ARPES band maps and momentum distribution curves (MDCs)
as a function of surface momentum 
along the  $\overline{\textrm M}$--$\overline{\Gamma}$--$\overline{\textrm M}$   direction
of the (001) surface Brillouin zone.
The soft x-ray photon energy was taken to be $h \nu =$ 380 eV
and the sputtering time was increased form $0$ s in panel (a)  to 112 s in panel (e).
 Additional SX-ARPES measurements with a photon energy of $h \nu =$ 725 eV are presented in Fig.~\ref{SFig1} of 
the Supplemental Material (SM)~\cite{supplement}.

\emph{Discussion.--} 
We start by discussing the spectra of the unsputtered sample, Figs.~\ref{fig:AllData}(a) and ~\ref{fig:AllData}(f). 
In the band map of Fig.~\ref{fig:AllData}(a) the bulk valence bands of the Bi-$p$ and Se-$p$ orbitals 
are clearly visible at $\sim$0.5~eV and $\sim$1.2~eV, respectively. 
Near the $\overline{\Gamma}$ point these valence bands exhibit a characteristic
``M-shape" dispersion, which agrees with ab initio DFT calculations~\cite{beniaPRB13,Zhang2009}. 
The Dirac surface state, however, is barely detectable, neither at $h \nu = 380$~eV nor at
$h \nu = 725$~eV~\cite{supplement}. This is due to the large amount of surface disorder introduced
by the adsorbates, which 
 smears out the quasiparticle peak of the Dirac state.
 Indeed, we find that
 the ARPES band map of Fig.~\ref{fig:AllData}(a) is in good agreement
 with the simulated spectral function of the Bi$_2$Se$_3$ 
 surface in the presence of strong Gaussian disorder [left column of Fig.~\ref{fig:Fig2}(b)].
 We note that the photon energies are chosen such to select a $k_z$ where the bulk conduction band is not observed.
 To our best knowledge this constitutes the first measurement of the surface state of a topological insulator at such high photon energies.

Sputtering the surface for a few seconds reduces   Gaussian disorder  by removing  adsorbates, but creates local defects, which increases  unitary disorder. As a net effect, we find that the quasiparticle peaks in the spectral function of the bulk and surface bands become considerably sharper [Figs.~\ref{mFig3}(b) and \ref{mFig3}(g)].
The ``V-shaped" dispersion and the Dirac point of the surface state, which is located at $\sim$0.3~eV, are now clearly discernible.
This observation corresponds well with the calculated spectral function of the middle column of Fig.~\ref{mFig2}(b), where
the sputtering process was simulated by adding 20\% surface vacancies.

For longer sputtering times the concentration of surface defects is further increased, but the Dirac surface state
remains visible, albeit with a { broader quasiparticle peak} [Figs.~\ref{mFig3}(c) and \ref{mFig3}(d)].
This observation can be explained by noting that a large density of defects eventually leads
to an insulating surface layer, thereby pushing the Dirac surface state to the second and third inward unit cells. 
Since SX-ARPES is less sensitive to the second and third unit cells,
the spectral function intensity of the Dirac state becomes weaker.
This interpretation is confirmed by our numerical simulations of Fig.~\ref{mFig2},
which show that for 40\% defect density most of the spectral weight of the Dirac state is
concentrated in the second and third { quintuplet} layers.

Finally, we find that 
sputtering for more than one minute  radically changes the 
topography of the Bi$_2$Se$_3$ surface. That is, the entire crystal surface is cracked up into multiple tilted domains with  sizes
comparable to the synchrotron beam spot (74$\times$30 $\mu$m). 
As a result,  the ARPES spectra contain contributions from several domains
with relative shifts in surface momenta, since 
the normal photoemission angle now sensitively depends on the incident position of the synchrotron light [Fig.~\ref{mFig3}(e)].
While a broad signature of the bulk bands 
can still be observed, the Dirac surface state is completely absent in the spectra of the strongly sputtered samples. 
This indicates that the surface of Bi$_2$Se$_3$ has been rendered topologically trivial up to a depth
of more than three unit cells, i.e., beyond the probing depth of SX-ARPES.

\emph{Conclusion.--}\label{Sec:Conc}
In conclusion, we have investigating how  
 topological insulator surface states are affected by disorder
caused by neon sputtering and surface adsorbates.
By comparing numerical simulations to SX-ARPES measurements of Bi$_2$Se$_3$,
we have shown that the surface adsorbates correspond to Gaussian disorder,
while the local defects introduced by sputtering represent
unitary scatterers. We have demonstrated that the effects 
of Gaussian disorder can be reduced by sputtering, which 
removes surface adsorbates but at the same time introduces
 defects and vacancies. Since the latter have a weaker effect 
than Gaussian disorder,
 the ARPES signal of Bi$_2$Se$_3$ surfaces
 that have been exposed to air (or UHV rest gas) become
 sharper upon sputtering. Our findings demonstrate the extreme
 robustness of the  Bi$_2$Se$_3$ surface state against
 any type of surface disorder, thereby confirming its topological origin.

\acknowledgments
This work was supported by the Swiss National Science Foundation Project No. PP00P2\_144742\ 1.
APS was supported in part by the US National Science Foundation under Grant No. NSF PHY11-25915.  PH, JLM and BBI acknowledge support by the VILLUM foundation, the DFG under SPPI666 and the Danish National Research Foundation.

\bibliographystyle{apsrev}
\bibliography{BiSe_refs}

\begin{thebibliography}{50}
\bibitem[{par()}]{parameters}
\bibinfo{note}{In our numerical calculation we set $A_1=0.540$ eV, $A_2=0.673$
  eV, $B_1=0.301$ eV, $B_2=4.55$ eV, $D_1=0.0391$ eV, $D_2=1.57$ eV, $M=0.28$
  eV, and $\mu=-0.0068$ eV.}
\end{thebibliography}


 \clearpage
\newpage

\appendix

\setcounter{figure}{0}
\makeatletter 
\renewcommand{\thefigure}{S\@arabic\c@figure} 

\makeatother

\begin{widetext}

\begin{center}
\textbf{
\large{Supplemental Material for}}
\vspace{0.4cm} 

\textbf{
\large{
``Sputtering induced re-emergence of the topological surface state in Bi$_2$Se$_3$" } 
}
\end{center}

\vspace{0.1cm}

\begin{center}
\textbf{Authors:} 
Raquel Queiroz
Gabriel Landolt
Stefan Muff
Bartosz Slomski
Thorsten Schmitt
Vladimir N. Strocov
Jianli Mi
Bo Brummerstedt Iversen
Philip Hofmann
J\"urg Osterwalder
Andreas P. Schnyder
J. Hugo Dil
\end{center}

\vspace{0.5cm}

{ We present in this supplemental material additional SX-ARPES measurements 
and give the definition of the density of states distribution.

\section{I.~~~Additional SX-ARPES spectra}

Figure~\ref{SFig1}  shows the SX-ARPES band maps and momentum distribution curves with a photon energy of $h \nu = 725$~eV.
This figure should be compared to the SX-ARPES spectra with  $h \nu = 380$~eV, which are depicted in Fig.~\ref{fig:AllData} of the main text. 
SX-ARPES with $h \nu = 725$~eV is sensitive to a different region of the bulk Brillouin zone than
SX-ARPES with $h \nu = 380$~eV. This is the reason why the bulk spectrum in Fig.~\ref{SFig1} is quite different 
from the one of Fig.~\ref{fig:AllData}. 
The Dirac surface state, however does not have any $k_z$ dispersion, and therefore its photoemission spectrum does not
depend on photon energy $h \nu$.
Moreover, we find that the dependence on surface  sputtering is qualitatively similar to the one  of Fig.~\ref{fig:AllData}.}

\begin{figure*}[h!]
\begin{center}
\includegraphics[width=0.85\textwidth]{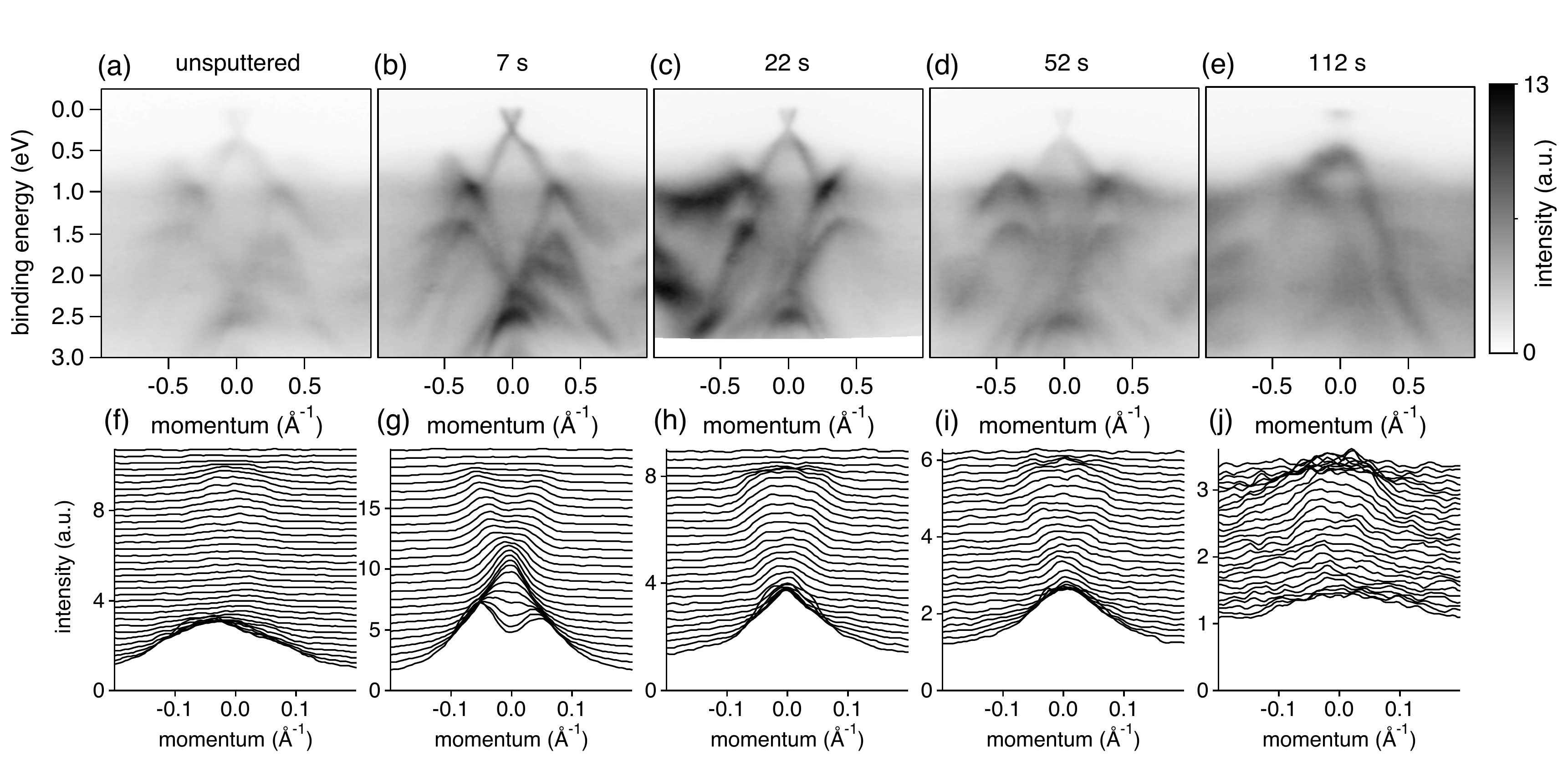}
\caption{
SX-ARPES measurements of the (001) Bi$_2$Se$_3$ surface with a photon energy of $h \nu = 725$~eV. 
The upper row [panels (a) to (e)] shows the band maps along the  $\overline{\textrm M}$--$\overline{\Gamma}$--$\overline{\textrm M}$ crystallographic direction for different sputtering times. 
The ARPES intensity is indicated by the grey scale, with black and white corresponding to high and low intensity, respectively. 
The lower row  [panels (f) to (j)] displays the momentum distribution curves (MDCs) for binding energies ranging form 0 meV to 442 meV with a spacing of 13 meV. For clarity each MDC is shifted vertically.
}
\label{SFig1}
\end{center}
\end{figure*}

\section{II.~~~Distribution of local density of states} 

{ Information about the localization or delocalization properties of the Dirac surface state can be obtained by computing
the  distribution of the local density of states (LDOS) $P [ \tilde{\rho}_l (\omega) ]$~{ \cite{schubert_vollhardt_PRB_10}.}
The probability distribution  $P [ \tilde{\rho}_l (\omega) ]$
is defined in terms of the normalized LDOS on the $l$th { quintuplet} layer
\begin{eqnarray}
 \tilde{\rho}_l (\omega) =  \tilde{\rho}_l (\omega) / \langle  \tilde{\rho}_l (\omega) \rangle ,
\end{eqnarray}
where $\langle  \tilde{\rho}_l (\omega) \rangle$ denotes the mean value of the LDOS.
The distributions shown in Fig.~\ref{fig:Fig2}(c) in the main text have been computed for an ensemble of
one thousand disordered Hamiltonians with periodic boundary conditions along the $x$ and $y$ directions.

We note that a distribution $P [ \tilde{\rho}_l (\omega) ]$ centered at $\tilde{\rho}_l (\omega)=1$ indicates that the majority of states are delocalized, 
while a distribution peaked at $\tilde{\rho}_l (\omega)=1$ signals that most of the states are localized.
The left (right) tails of the distribution give us insight, whether 
there exists a minority of localized (delocalized) states.
To quantify the size of these tails we introduce the left and right standard deviations, $\sigma_L$ and $\sigma_R$,
of $P [ \tilde{\rho}_l  ]$, which are given by
\begin{eqnarray}
\sigma^2_L
&=&
\left(
2 \sum_{\tilde{\rho}_{l} = 0}^{\rho_{\textrm{max}}}
P [ \tilde{\rho}_{l} ] - P [ \rho_{\textrm{max}} ] 
\right)^{-1}
 \sum_{\tilde{\rho}_{l} = 0}^{\rho_{\textrm{max}}}
2~P [ \tilde{\rho}_{l} ]\left( \tilde{\rho}_{l}-\rho_\textrm{max}\right)^2 
 ,
\\
\sigma^2_R
&=&
\left(
2 \sum_{\tilde{\rho}_{l} = \rho_{\textrm{max}}}^\infty
P [ \tilde{\rho}_{l} ] - P [ \rho_{\textrm{max}} ] 
\right)^{-1}
 \sum_{\tilde{\rho}_{l} = \rho_{\textrm{max}}}^\infty
2~P [ \tilde{\rho}_{l} ]\left( \tilde{\rho}_{l}-\rho_\textrm{max}\right)^2 .
\end{eqnarray}
In the above definitions we have treated 
 $\tilde{\rho}_l  > \rho_{\textrm{max}}$ and $\tilde{\rho}_l   < \rho_{\textrm{max}}$ as independent distributions, 
where $\rho_{\textrm{max}}$ is the maximum or the distribution $P[ \tilde{\rho}_l  ]$. The numerical results were obtained for 1000 disorder configurations of three dimensional lattices in real space, with $25^3$ lattice sites and periodic boundary conditions along the $(x,y)$ directions. }

\end{widetext}

 \clearpage

\end{document}